\newcommand{\be}{\begin{equation}}
\newcommand{\ee}{\end{equation}}
\newcommand{\bea}{\begin{eqnarray}}
\newcommand{\eea}{\end{eqnarray}}
\newcommand{\nn}{\nonumber}

\input epsf.sty

\documentclass[12pt]{iopart}

\begin{document}

\title[Comment on the paper `An analytic functional form...', PPCF {\bf 55} (2013) 095009]
{Comment on the paper `An analytic functional form for
characterization and generation of axisymmetric plasma boundaries', PPCF {\bf 55} (2013) 095009}

\author{Ap Kuiroukidis$^{1}$ and G. N. Throumoulopoulos$^{2}$}

\address{$^{1}$Technological Education Institute of Serres 621 24 Serres, Greece
\\
$^{2}$Physics Department, University of Ioannina,
 GR 451 10 Ioannina, Greece}

\ead{kouirouki@astro.auth.gr,$\; \; $gthroum@uoi.gr }



In Ref. \cite{luce} it was proposed an analytic form to describe the boundary of an axisymmetric plasma. The form determines analytically smooth boundaries, e.g. D-shaped ones, and  numerically boundaries having an X-point. 
Here we propose  an alternative form describing boundaries with X-points fully analytically.  

The two
coordinates are $(R,z)$ and they are normalized with respect to the major radius $R_{0}$.
For example for ITER we have $R_{0}=6.2m$.  The boundary  is up-down asymmetric and it consists of
a smooth upper part and a lower part possessing an  X-point.
For the upper part we have four parameters. One is the inverse aspect ratio
$\epsilon_{0}=a/R_{0}$. For example in the case of ITER it is $a=2.1m$ and thus
$\epsilon_{0}=0.338$. The other three are the upper elongation $\kappa$,  the upper
triangularity $\delta$ and a parameter $n$  related to the steepness of the triangularity of
the upper boundary curve, i.e. to the mean value of the derivative
 $d\rho/d\theta$ [see Eqs. (\ref{upper}) and (\ref{taf}) below]. For the ITER case it is  $\kappa=1.86$ and  $\delta=0.5$. Introducing the  normalized coordinates
$\rho=R/R_{0}$ and $\zeta=z/R_{0}$,  we have for the $\zeta$ coordinate of the uppermost point (see Fig. 1): $\zeta_{u}=\kappa\epsilon_{0}$ and
$\delta=(1-\rho_{\delta})/\epsilon_{0}$. 
\begin{figure}[ht!]
\centerline{\mbox {\epsfxsize=12.5cm \epsfysize=10.cm \epsfbox{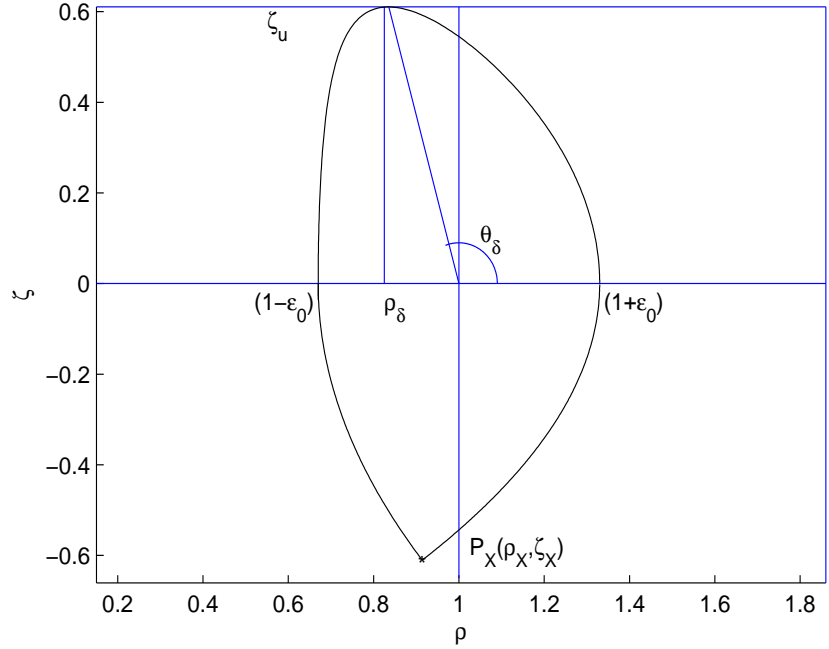}}}
\caption[]{The boundary of an axisymmetric plasma having a saddle point, $P_X$,  as determined by Eqs. (\ref{upper}), (\ref{ll}) and (\ref{lr}) containing    six shaping parameters.}
\label{fig1}
\end{figure}

The form for the upper part of the
boundary is given by
\bea
\label{upper}
\rho&=&1+\epsilon_{0}cos(\tau+\alpha sin(\tau))\nn \\
\zeta&=&\zeta_{u}sin(\tau)
\eea
where $\alpha=sin^{-1}(\delta)$. Thus the following relations hold:
$\rho_{\delta}=1-\delta\epsilon_{0}$ and \\
$\theta_{\delta}=\pi-tan^{-1}(\kappa/\delta)$. The parameter $\tau$ is any increasing
function of the usual polar angle $\theta$, satisfying $\tau(0)=0$, $\tau(\pi)=\pi$
and $\tau(\theta_{\delta})=\pi/2$. In our form we have taken
\bea
\label{taf}
\tau(\theta)&=&t_{0}\theta^{2}+t_{1}\theta^{n}\nn \\
t_{0}&=&\frac{\theta_{\delta}^{n}-\frac{1}{2}\pi^{n}}{\pi\theta_{\delta}^{n}-\theta_{\delta}^{2}\pi^{n-1}}\nn
\\
t_{1}&=&\frac{-\theta_{\delta}^{2}+\frac{1}{2}\pi^{2}}{\pi\theta_{\delta}^{n}-\theta_{\delta}^{2}\pi^{n-1}}
\eea
The uppermost point of the bounding curve has coordinates $(\rho_{\delta},\zeta_{u})$.

Now we consider the lower part of the boundary  which is located
at the $\zeta<0$-part of the plane. It depends on
two additional parameters, namely the lower elongation ${\bar \kappa}$ and
the lower triangularity ${\bar \delta}$. In terms of these parameters we have
$\zeta_{d}={\bar \kappa}\epsilon_{0}$ and
${\bar \theta}_{\delta}=\pi-tan^{-1}({\bar \kappa}/{\bar \delta})$, where $-\zeta_{d}$ corresponds to the X-point. 
The equation for the left-lower part of the bounding curve, with respect to the  X-point, is given by
\bea
\label{ll}
\rho_{l}&=&1+\epsilon_{0}cos(\theta)\nn \\
\zeta_{l}&=&-[2p_{1}\epsilon_{0}(1+cos(\theta)]^{1/2}\nn \\
p_{1}&=&\frac{\zeta_{d}^{2}}{2\epsilon_{0}(1+cos{\bar \theta}_{\delta})},\; \; \;
(\pi\leq \theta\leq 2\pi-{\bar \theta}_{\delta})
\eea
while that for the right-lower part is given by
\bea
\label{lr}
\rho_{r}&=&1+\epsilon_{0}cos(\theta)\nn \\
\zeta_{r}&=&-[2p_{2}\epsilon_{0}(1-cos(\theta)]^{1/2}\nn \\
p_{2}&=&\frac{\zeta_{d}^{2}}{2\epsilon_{0}(1-cos{\bar \theta}_{\delta})},\; \; \;
(2\pi-{\bar \theta}_{\delta}\leq \theta\leq 2\pi)
\eea
The coordinates of the X-point are
$(\rho_{X},\zeta_{X})=(1+\epsilon_{0}cos{\bar \theta}_{\delta},-\zeta_{d})$.
 At the midplane $\zeta=0$ the upper and
the right- and left-lower parts of the boundary meet smoothly, that is
 the derivative $d\zeta/d\rho$ is continuous there.

In summary, alternatively to the  numerical functional-form determination of the boundary of an axisymmetric plasma having an X-point proposed in Ref. \cite{luce},  we propose here a completely analytic form for boundaries of this kind.  This form contains six free parameters associated with the boundary shaping and,   as the case of Ref. \cite{luce},   should be useful for  studies of fusion plasmas.



\section*{Aknowledgments}\

This work was supported by (a) the National Programme for the Controlled Thermonuclear Fusion, Hellenic Republic, (b) the European Union's Horizon 
2020 research and innovation programme under grant agreement number 
633053. The views and opinions expressed herein do not necessarily 
reflect those of the European Commission.



\section*{References}

\begin{thebibliography}{0}

\bibitem{luce}T C Luce, Plasma Phys. Control. Fusion {\bf 55}
(2013) 095009; idid {\bf 55}  (2013) 119501 (corrigendum).


\end{thebibliography}
\end{document}